\documentclass[12pt,a4paper]{article}
\newif\ifpdf\ifx\pdfoutput\undefined\pdffalse\else\pdfoutput=1\pdftrue\fi
\usepackage{graphicx}
\usepackage{t1enc}
\usepackage[latin1]{inputenc}
\usepackage[english]{babel}
\begin{document}
%\pdfgraphics 

\title{Simple queueing approach to segregation dynamics in Schelling model}
\author{Pawel Sobkowicz}
\maketitle

\begin{abstract}{A simple queueing approach for segregation of agents in modified one dimensional Schelling segregation model is presented.
The goal is to arrive at simple formula for the number of unhappy agents remaining after the segregation.}
\end{abstract}
The importance of the original Schelling model of segregation was mainly due to the combination of simplicity and ability to reproduce phenomena observed in sociology. Particular success of the model was related to description of residential segregation. Whether for a street or a city, the simple 1D or 2D (checkerboard) geometries were reasonable representations of real situation. And the results have proven to enhance our understanding of the social behaviour.

It is worth noting that since 1990's considerable amount of literature pointed out that there are many aspects of social phenomena which may be described in dynamical agent models, but which require going beyond simple nearest neighbour 1D and 2D models. Several phenomena, for example clique formation and clustering are much better represented in the network based approach. Network models allow also much greater flexibility in describing interactions between agents, in particular showing fat-tailed distribution of the number of neighbours and `small world' phenomena. In general, models based on  nearest neighbour  1D and 2D geometries can provide only limited range of description  of real life  phenomena, where interactions between agents are usually long range and structureless.

However, even such limited models may provide a glimpse of understanding of the mechanisms and processes. With this goal in mind, this paper presents a very simple approach to dynamics of segregation of agents in 1D, short interaction case.

The model is based on assumptions similar to those published in \cite{dallastaarxiv}. We consider a one dimensional chain of $L$ sites and $N$ agents ($L>N$, both being large), who can be in two states, $+$ and $-$. Some of the sites in a chain are vacant. The densities of the $+$, $-$ and vacant sites are considered to be constant and denoted, respectively, by $\rho_+$, $\rho_-$ and $\rho_0$.
To stress the similarity of the model to spin dynamics it is useful to introduce another quantity, corresponding to overall magnetization density, defined as  $m=\rho_+-\rho_-$.

An agent is considered to be unhappy when certain fraction $f^*$ of its neighbours is of different type. We can model in this fashion the reactions of agents to `unfriendly' neighbourhood. In the simplest model, considered here, we look only to the nearest neighbours as influencing the agent's behaviour. This behaviour is simple: an agent may try to improve its situation by moving to a vacancy that would improve (or at least not worsen) its situation. Starting from a random distribution of the agents the subsequent jumps should lead to Schelling segregation of the population.

We shall further limit ourselves to the simple case of the `constrained model' where only unhappy agents are motivated to move. The `unconstrained' situation, when also happy agents move as long as this does not make them unhappy, introduced by \cite{dallastaarxiv} is not considered here. The basic quantity we are interested in here is the number of unhappy agents at the end of the process of jumps, $u_{\infty}$. Instead of random selection of the agents considering whether to jump or not we shall use a simple queueing model, in which agents play a sort of `musical chairs' game. We assume an `intelligent', God's-eye view algorithm designed to maximize the results of the optimization process. Such model of segregation presented here is rather close to the original Schelling algorithm, which has assumed that the unhappy agents would be queued up and attempt to improve their situation in ordered fashion.

In the following we will use, as far as possible, the notation of \cite{dallastaarxiv}.
Let's assume without decreasing the generality of the treatment that $m\ge0$, i.e. $\rho_+ \ge \rho_-$. 
We would thus call $+$ agents the majority agents, and the $-$ would be the minority.

For the 1D nearest neighbour model a given agent may have 0, 1 or 2 agents as neighbours. As noted before, $f^*=1/2$ condition means  that an agent is unhappy only if it is surrounded by two opposite agents. Let us consider this value of $f^*$ first.

 For  a random initial configuration of agents we can calculate probabilities of $+$ and $-$ agents to be in unhappy state. 
\begin{eqnarray}
\rho_+^U & = & \rho_+ \rho_-^2 \\
\rho_-^U & = & \rho_+^2 \rho_- 
\end{eqnarray}
This can be expressed through $m$ and $\rho_0$:
\begin{eqnarray}
\rho_+^U & = & \left((1-\rho_0)^2 - m^2\right)\left( 1-\rho_0 -m \right)/8 \\
\rho_-^U & = & \left((1-\rho_0)^2 - m^2\right)\left( 1-\rho_0 +m \right)/8 
\end{eqnarray}
It is worth noting that the average initial number of unhappy minority ($-$) agents is greater than the number of unhappy majority agents.

Starting from such a random configuration let us now try to estimate the outcome of the dynamic process, in which unhappy agents improve their utility function. In the constrained model \textbf{only} unhappy agents are allowed to move, and only if the move improves their utility function.

Let's describe here the queueing approach. Let us assume that somewhere in the system there is an initial vacancy `friendly' for a $+$ agent. All that is needed at this stage is a single such vacancy. (It is worth noting that in the original Schelling model, and agent was simply inserted between two other agents, regardless of any formal vacancies). We move an unhappy $+$ to this vacancy. But for a $+$ to be unhappy it must have come from a $-+-$ environment. And thus it leaves behind a vacancy ($-0-$), which is for sure `friendly' 
for an unhappy $-$ agent (from a $+-+$ environment). The next step is then migration of such unhappy $-$ agent, which, in turn leaves a $+$ friendly vacancy ($+0+$). The process would be repeated until we run out of the suitable candidates for swapping, i.e. it would swap (into happiness state) $\rho_+^U$ $+$ and $-$ agents. It is worth noting that this process is feasible even if there is just one suitable initial vacancy. In such case, even a random process would unerringly pick the right `targets'. 

The remaining unhappy minority agents, $\rho_-^U-\rho_+^U$ could improve their situation if there are suitable vacancies in the system. The number of such `$-$' friendly vacancies is
\begin{equation}
\rho_V^{F-} = \rho_0 ( 1 - \rho_+^2) = \rho_0 \left( 1 - (1-\rho_0+m)^2/4 \right)
\end{equation}

The final amount of unhappy agents (all of minority type) would thus be, on average,
\begin{equation}
u_{\infty} = \rho_-^U - \rho_+^U - \rho_V^{F-}
\end{equation}

Figure 1 shows the behaviour of $u_{\infty}$ for some values of $m$ as function of the vacancy density $\rho_0$. 
It is worth noting that with the queueing method, if $m=0$ and $\rho_0 \rightarrow 0$ the amount of unhappy agents for a specific starting condition would be proportional to statistical difference between initial distribution of unhappy $+$ and $-$ agents. This would generally be proportional to $\sqrt{L}$, so in the limit of large $L$ should become negligible. The average value $u_{\infty}$ for many initial configurations for $m=0$ is zero, as it is in the Dall'Asta et. al. unconstrained model. 

In the $f^*=1/2$ case we can compare the results of the queueing model with Figure~1 of the Dall'Asta et. al. paper \cite{dallastaarxiv}. The results are qualitatively similar, for example they show the same gradual decrease of the density of unhappy sites with increasing $\rho_0$ and `perfect happiness' above some threshold value $\rho_0^*(m)$. The absolute values are, however bigger roughly by a factor of 2, which reflects the simpleness of the current model (for example the fact that one transition may make more than one agent happy, for example moving out the central $+$ agent from the environment
$(+-+-+)$ makes, simultaneously, both neighbouring $-$ agents happy. This should further decrease the amount of unhappy agents $u_{\infty}$). Such considerations were neglected in present reasoning for the sake of simplicity.

We now turn to situation of $f^* < 1/2$ when the process changes profoundly. 
Now an agent is happy only if it is surrounded by the same type agents. Thus, in the constrained model migration is possible only to vacancies that do not have any opposite agent in the neighbourhood. For example, to migrate a $+$ one needs a vacancy of three following types: $+0+$ or $00+$ or $+00$. All other vacancies are unsuitable. The ability to successfully migrate is thus much lower. 

When we start, as before, from a random 1D chain we can calculate the initial number of unhappy agents
\begin{eqnarray}
\rho_+^U & = & \rho_+ \left(1-(\rho_+ +  \rho_0)^2  \right) \\
\rho_-^U & = & \rho_- \left(1-(\rho_- +  \rho_0)^2  \right)
\end{eqnarray}
or
\begin{eqnarray}
\rho_+^U & = & \left(1- (1+\rho_0+m)^2/4 \right)\left( 1-\rho_0 +m \right)/2 \\
\rho_-^U & = & \left(1- (1+\rho_0-m)^2/4 \right)\left( 1-\rho_0 -m \right)/2 
\end{eqnarray} 

It is much more difficult to build a successful queueing system now, as it is no longer true that a successfully migrating unhappy agent would leave behind an environment ready for accepting the opposite agent. In fact only in $-+0$, $0+-$ and $-+-$ would migration of the $+$ create a vacancy ready to accept a $-$. Let's call such positioning `optimal'.

The queueing procedure would then be divided into three steps. First, we would use a single `starting vacancy' to make the chain of exchanges, but only between the optimally placed agents. The densities of such optimally placed agents is different for $+$ and $-$ and is given by
\begin{eqnarray}
\rho_+^s & = & \rho_- \left(2 \rho_0\rho_+ +\rho_+^2  \right) \\
\rho_-^s & = & \rho_+ \left(2 \rho_0\rho_- +\rho_-^2  \right)
\end{eqnarray}
where $s$ stands for `swap friendly' density, for example $\rho_+^s$ is the density of $-$ sites that would leave a friendly vacancy for a $+$. It is worth noting that $\rho_+^s > \rho_-^s$. As the exchanges go one by one, only $\rho_-^s$ of them can be made. This step reduces the number of unhappy sites to
\begin{eqnarray}
\rho_+^U & \rightarrow & \rho_+^U - \rho_-^s \\
\rho_-^U & \rightarrow & \rho_-^U - \rho_-^s
\end{eqnarray}

The next step is to look for vacancies already present in the system that could accommodate, in a single transfer, an unhappy $+$ or $-$. Such vacancies are, respectively, given by $+0+, +00, 00+$ and $-0-, -00, 00-$ configurations. Their average densities are given by
\begin{eqnarray}
\rho_V^+ & = & \rho_0 \left(2 \rho_0\rho_+ +\rho_+^2  \right) \\
\rho_V^- & = & \rho_0 \left(2 \rho_0\rho_- +\rho_-^2  \right)
\end{eqnarray}

The remaining (if any) unhappy agents of any type can then, at the last stage, be moved to totally empty environments ($000$), which have average density of $\rho_{VVV} = \rho_0^3$. Thus the final density of unhappy sites is
\begin{equation}
u_{\infty} = \rho_+^U + \rho_-^U - 2 \rho_-^s - \rho_V^+ - \rho_V^- -\rho_{VVV}
\end{equation}

Figure 3 shows examples of $u_{\infty}$ for some values of $m$. In the $f^*<1/2$ model, the average number of unhappy agents at $\rho_0 \rightarrow 0$ is $1/2$. Figure 4 shows dependence of the $u_{\infty}$ on $m$ for a small value of $\rho_0$.

\begin{figure}[p]
	\centering
		\includegraphics[scale=0.8]{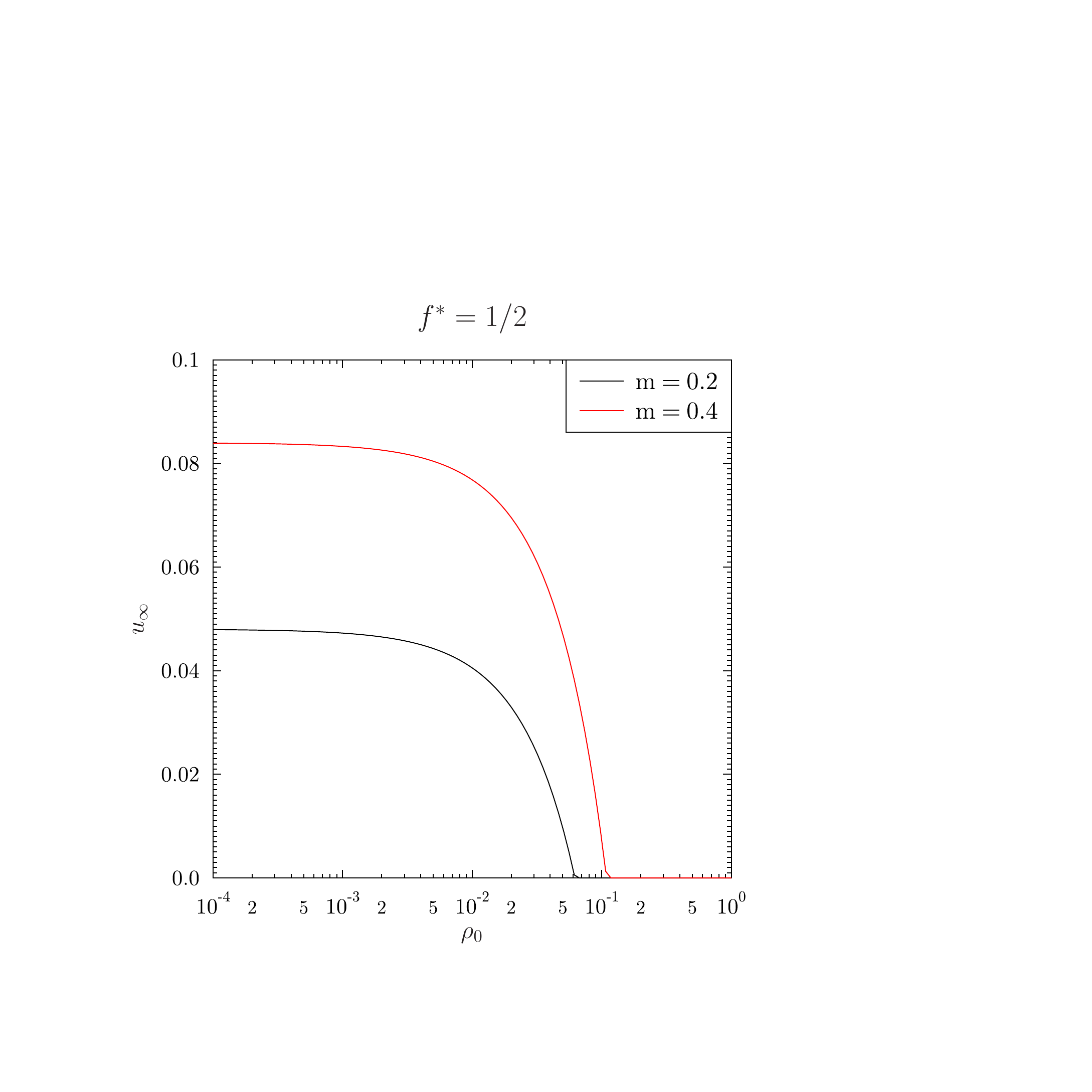}
	\caption{$u_{\infty}$ as function of $\rho_0$ for several values of $m$ for the $f^*=1/2$ case.}
	\label{fig:easyhappy}
\end{figure}

\begin{figure}[p]
	\centering
		\includegraphics[scale=0.8]{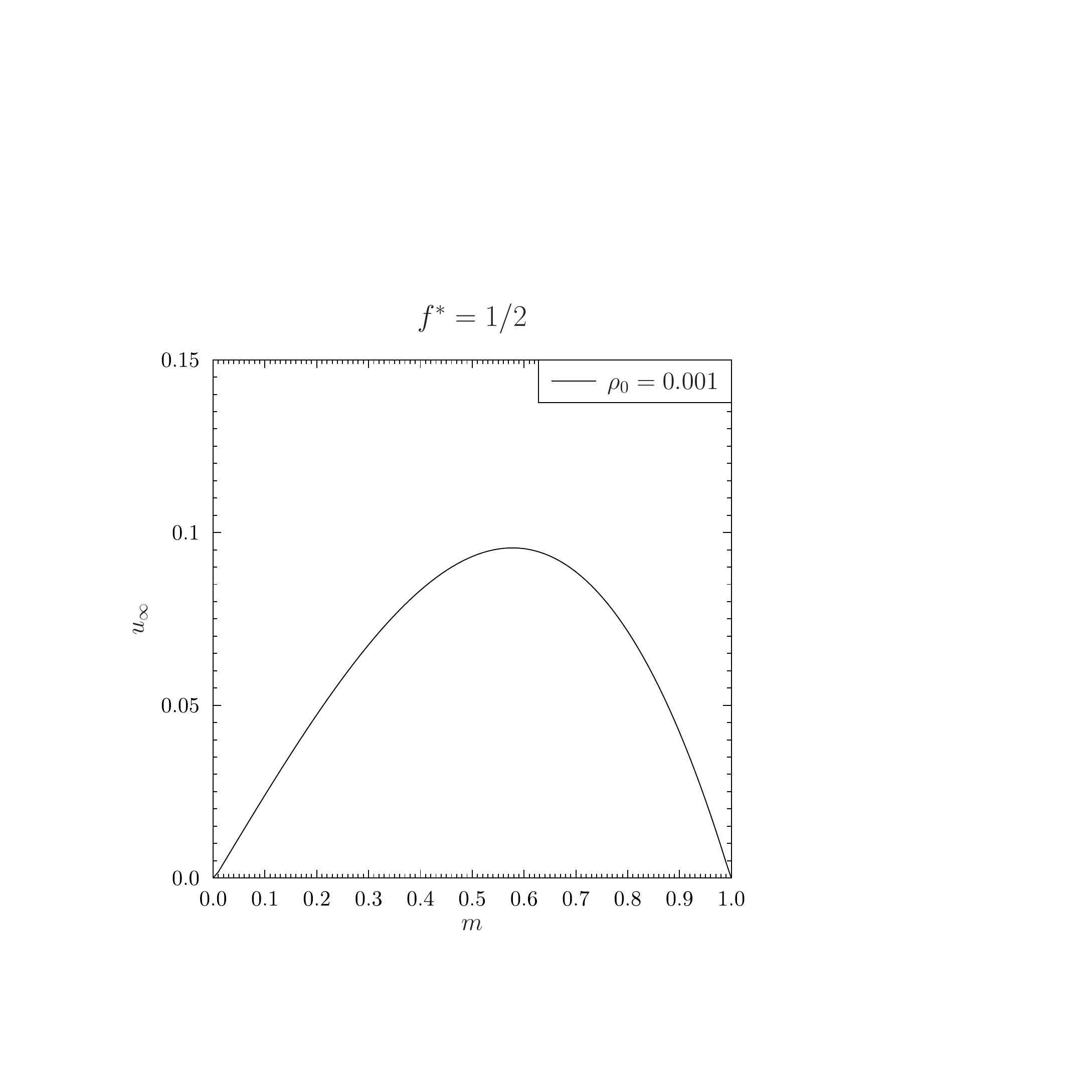}
	\caption{$u_{\infty}$ as function of $m$ for small value of $\rho_0 = 0.001$ for the $f^*=1/2$ case.}
	\label{fig:easyhappy-m}
\end{figure} 
\vspace{0.5cm}

\begin{figure}[p]
	\centering
		\includegraphics[scale=0.8]{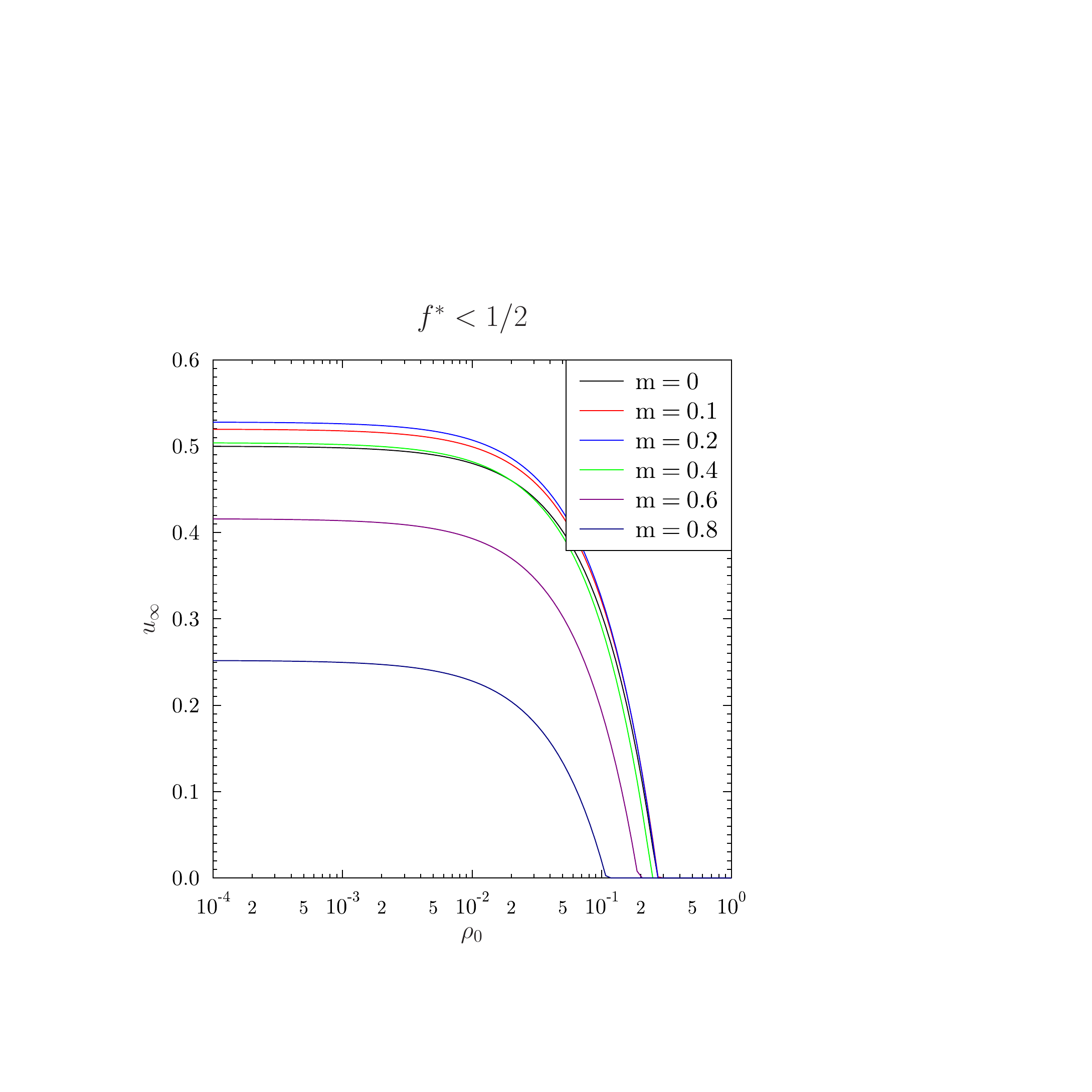}
	\caption{$u_{\infty}$ as function of $\rho_0$ for several values of $m$ for the $f^*<1/2$ case.}
	\label{fig:hardhappy}
\end{figure}

\begin{figure}[p]
	\centering
		\includegraphics[scale=0.8]{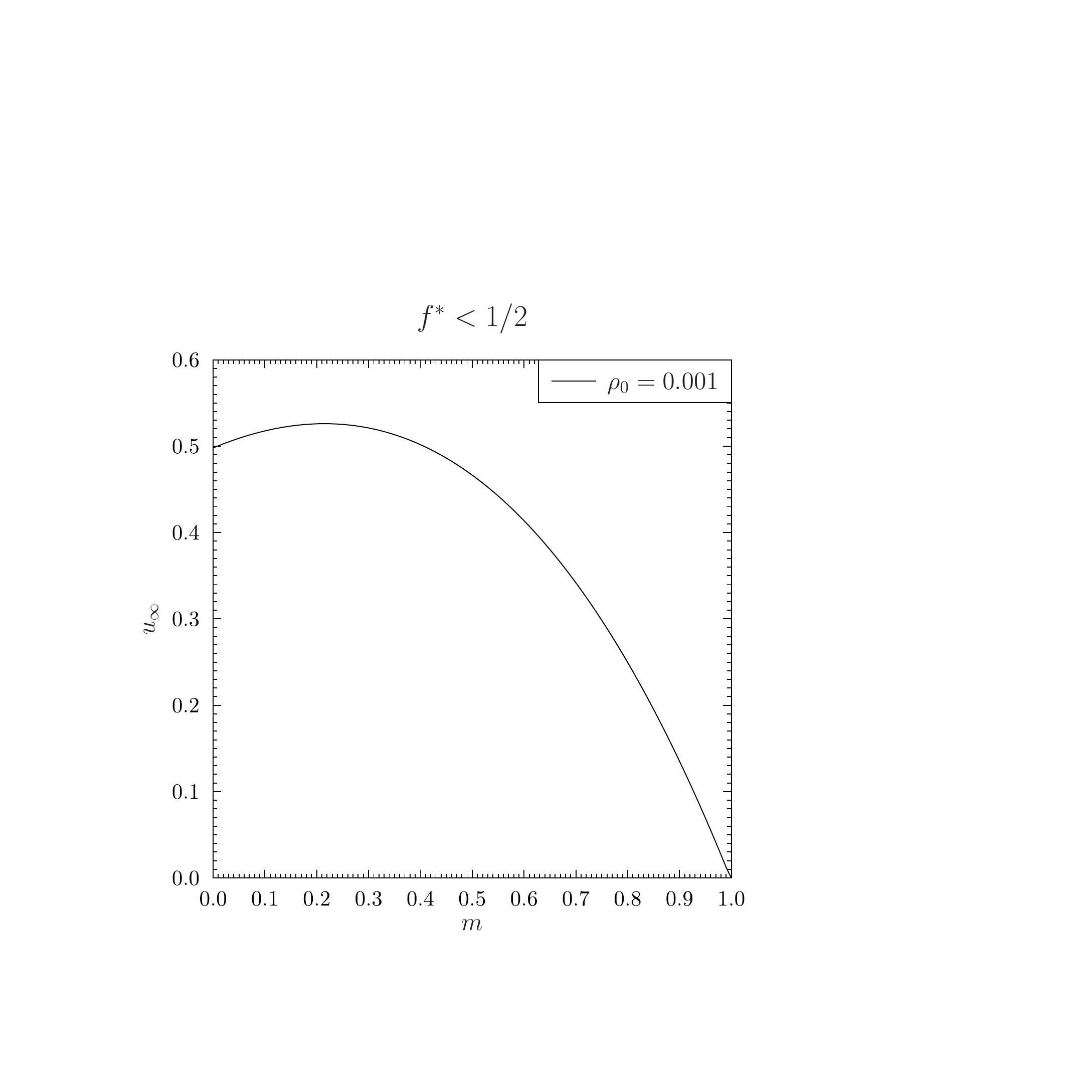}
	\caption{$u_{\infty}$ as function of $m$ for small value of $\rho_0 = 0.001$ for the $f^*<1/2$ case.}
	\label{fig:hardhappy-m}
\end{figure} 
\vspace{0.5cm}

\bibliography{dallastabib}

\end{document}